\documentstyle[twocolumn,aps,epsfig]{revtex}
\begin{document}
\draft
\title{ FIRST-ORDER TRANSITION OF TETHERED MEMBRANES IN 3D SPACE}
\author{J.-Ph. Kownacki and H. T. Diep}
\address{Laboratoire de Physique Th\'eorique et Mod\'elisation,
CNRS-Universit\'e de Cergy-Pontoise, UMR 8089\\
5, mail Gay-Lussac, Neuville sur Oise, 95031 Cergy-Pontoise Cedex, France\\
E-mail: diep@ptm.u-cergy.fr}
\maketitle
\begin{abstract}
We study  a model of phantom tethered membranes, 
embedded in three-dimensional space, by extensive Monte Carlo
simulations. The membranes have hexagonal lattice structure where each
monomer is interacting with six nearest-neighbors (NN).  Tethering
interaction between
NN, as well as curvature penalty between NN triangles are taken into account. 
This model is new in the sense that  
NN interactions are taken into account by a truncated Lennard-Jones
(LJ) potential including both repulsive and attractive parts.

The main result of our  study is that the system undergoes a {\it first
order  
crumpling transition} from low temperature flat phase to
high temperature crumpled phase, in contrast
with early numerical results on models of tethered membranes.

\end{abstract}
\vspace{1cm}
\vspace{1cm}

\section{Introduction}
Statistical mechanics of membranes is a rich
subject and has been studied since about twenty years. 
Motivations to obtain a full understanding of the
behavior of these complex systems are enforced by many
experimental realizations. For recent reviews,
see \cite{bowi00} and \cite{wies00}. See also  \cite{jeru} for many 
introducing and pedagogical courses on the subject.

Membranes are two-dimensional fluctuating systems of monomers. 
According to their physical properties, membranes can be ``fluid''
or ``tethered''.
Fluid membranes consist of freely moving monomers, $i.e$ with Hamiltonian 
depending only on the shape of membranes. On the contrary, 
monomers in tethered membranes are
tied together by a tethering potential and their connectivity 
is fixed. In addition, a membrane can be self-avoiding if intersections 
with itself are forbidden. Otherwise, it is a phantom
membrane. In this paper, we focus our attention on a model of
tethered membranes with external curvature energy 
without self-avoidance.

Any realistic model should include self-avoiding interactions.
But phase diagrams of phantom membranes are rich
and contribute to understand the behavior of self-avoiding
membranes \cite{davi95,kard88.1}. It is now firmly established that
phantom membranes undergo crumpling transition between a flat and a
crumpled phase. Flat phase possesses long-range
orientational order between normals to the surfaces whereas crumpled phase is
totally disordered. However, the nature of the crumpling transition
is still puzzling.
Renormalization group (RG) calculations \cite{pacz88} with a Landau
continuous model
\cite{kard88.2}
predict a discontinuous phase transition when the dimension $d$ of
the embedding
space is lower than 219, including the physical case $d=3$. On the other
hand,  numerical simulations of lattice models
\cite{kant86,kant87,bowi96,whea93,whea96,baig89,baig94,ambj89,renk90,harn91,espr96}, 
large $d$ expansion \cite{davi88} and
calculations based on truncations \cite{ledo92} of the Schwinger-Dyson
equations are consistent with a continuous phase transition.

In this paper, we try to shed a light on this contradiction with
an extensive Monte Carlo (MC) study
on a model of
phantom tethered membranes with bending rigidity.
As it turns out,
our results show that the crumpling transition within our model is of
first order in agreement with the RG prediction.

Section II is devoted to a description of the model. Our method is described
in Section III and the results are shown in Section IV. Concluding remarks
are given in the last section.

\section{The model}
We consider a 2D lattice of monomers connected
in hexagonal structure, and embedded in the physical 3d euclidean space. 
The tethering potential between nearest-neighbor (NN) monomers
is a truncated Lennard-Jones
(LJ) potential. The curvature energy is a standard normal-normal
interaction between
NN triangles.
The distance between NN monomers are not allowed to be larger than an
upper bound distance $R_{max}$. Otherwise,
in absence of $R_{max}$,
monomers are no longer effectively
tethered at high temperature and the system becomes a gaz. In order to keep 
essential features of the LJ potential,  $R_{max}$ must be sufficiently
larger than $r_{0}$, the NN distance corresponding to
the minimum of the potential, so that $R_{max}$ lies
in the flat asymptotic region of the potential. However, to have an
actual tethered membrane,
$R_{max}$ should not be too large as discussed above.
Moreover, equilibration times increase as
$R_{max}$ increases, since NN distances are then allowed to grow more
and more.

The system is described by the Hamiltonian
\begin{equation}
{\cal H}=\sum_{<ij>} U(r_{\mbox{\tiny \it ij}}) -K
\sum_{<\alpha \beta>}{\bf n}_{\alpha}.{\bf n}_{\beta}
                             \label{equ.hami}
\end{equation}

The first sum is performed on pairs of NN monomers $<i,j>$ only,
and the second one is restricted to pairs of NN
triangles $<\alpha , \beta>$.
Tethering interaction between NN monomers labeled by $i$ and $j$
depends only on their distance
$r_{ij}$ in the 3d embedding space and is described by 
$U(r_{\mbox{\tiny \it ij}})$,
\begin{eqnarray}
U(r_{\mbox{\tiny \it ij}})&=& U_o\Big[\Big(\frac{r_{o}}{r_{\mbox{\tiny \it
ij}}}\Big)^{12}-
                2 \Big(\frac{r_{o}}{r_{\mbox{\tiny \it ij}}}\Big)^{6}\Big] \ \
\  
               \mbox{if $r_{\mbox{\tiny \it ij}} < R_{max}$}  \nonumber \\
         &=& 0 \mbox{\phantom{.}\hskip 3.55cm if $r_{\mbox{\tiny \it ij}} \ge
R_{max}$}
\end{eqnarray}
 with $r_{\mbox{\tiny \it ij}}=\| {\bf r}_i - {\bf r}_j \|$,
 ${\bf r_i}$ and ${\bf r_j}$
being the position vectors in the 3d space.
$r_{o}$ is the equilibrium 
distance between NN monomers. The second term in (\ref{equ.hami}) is
the external curvature
energy, with $K$ the bending rigidity.
The 3d-vector ${\bf n}_{\alpha}$ is defined as the normal unit vector of the
$\alpha -th$ triangle formed by three NN monomers.
Note that ${\bf n}_{\alpha}$ is defined for a counterclockwise
oriented triangle.

The phase space of the model depends on three parameters. We fix two of
them, namely
($U_o$ and $K$), and look for temperature-dependent properties.

\section{Numerical method}

We consider a membrane of linear size $L$.  The total number
of monomers is $N=L\times L$.
We choose $U_o=3$ and $K=1$. A more extensive study would require to explore
the complete phase space, but these particular values already give
interesting results.
The NN distance in the ground state is taken to be the unit
of distance, i.e. $r_o = 1$, and
the upper bound
$R_{max} = 4$. We use free boundary conditions,
at constant pressure in our simulations.
 
The following algorithm was used. Starting from the ground state where
monomers
are on the hexagonal  lattice sites, we heat the system to
a temperature $T$. We
equilibrate the system at variable volume.  The local
equilibration is done as follows:
we take a monomer and move it to  a nearby random position
in a cubic box of volume $\delta^3$ around its position, in the 3d space. 
This position is accepted if it lowers the  energy. Otherwise it is accepted 
with a probability according to the Metropolis algorithm. We repeat
this for all monomers:
we say we achieve one MC step/monomer. We choose $\delta \simeq 0.1$
to have an
acceptance of the order of 50 $\%$.

We define the following physical quantities -  
averaged total energy $\langle E\rangle$, averaged normal vector $<n>$, 
averaged NN distance $<d>$,  radius of gyration $R_g$ - 
with the following standard definitions  
\begin{equation}
\langle E\rangle=\langle\cal{H}\rangle
\end{equation}
\begin{equation}
<n>= \frac{1}{2(L-1)^2}  <|\sum_{\alpha} {\bf n}_{\alpha}|>
\end{equation}
\begin{equation}
<d>=\frac{1}{(3L-1)(L-1)} \sum_{<i,j>} <r_{ij}>
\end{equation}
\begin{equation}
R_g^2=\frac{1}{2 L^4} \sum_{i,j}<({\bf r}_i-{\bf r}_j)^2>
\end{equation}
where $<...>$ indicates thermal average and
the sum in $<d>$ is performed only
on NN links.

All the results described below are obtained after thermalization
of the system. This
requires about $10^6-10^7$ MC steps/monomer, depending on the
temperature. After thermalization,
measures are done on $10^{6}$-$10^7$ MC steps/monomer, depending
also on the temperature.

Error bars are calculated using a standard jack-knife algorithm.

\section{Results and analysis}
At low temperature, equilibrium configurations are configurations
of minimum energy.
This means that NN distance must be close to $r_o$, the minimum
of the LJ potential,
and that NN normals must be parallel to minimize curvature energy.
So, equilibrium
states correspond to a flat state, with $<n> \simeq 1$ and
$R_{g} \simeq L$. It
is an ordered phase.

At high temperature, maximal entropy configurations correspond to
crumpled states, where
the membrane is compact and occupies a very small volume in the
embedding space. It is a
completely disordered phase, with $<n>=0$ and $R_{g} \ll L$.

Between low and high temperatures, a crumpling phase transition
is expected. In this work,
we  study the nature of this transition to see
whether it is continuous or not.

\begin{figure}[tb] 
\epsfig{file=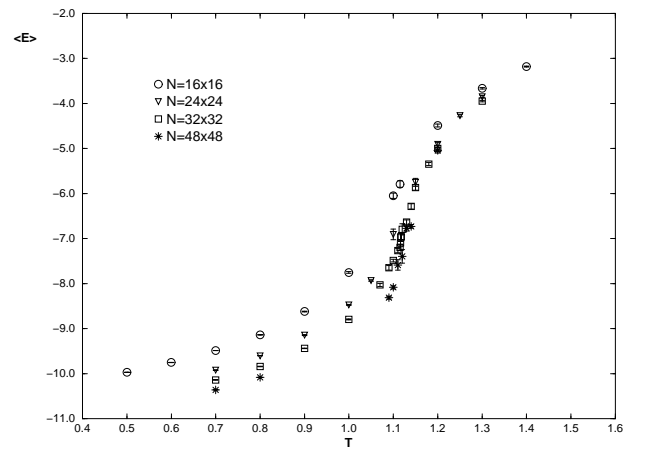,width=6.cm,angle=-90}
\vskip 0.5 truecm
\caption{\small{Averaged energy versus temperature, for $N=16\times 16$,
$24\times 24$, $32\times 32$ and $48\times 48$}}
\label{ener}
\end{figure}

\begin{figure}[htb] 
\epsfig{file=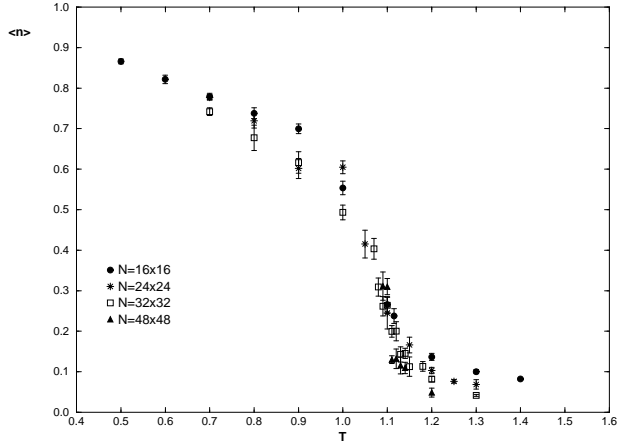,width=6.cm,angle=-90}
\vskip 0.5 truecm
\caption{\small{$<n>$ versus $T$, for
$N=16\times 16$, $24\times 24$, $32\times 32$ and $48\times 48$}}
\label{norm}
\end{figure}

\begin{figure}[htb] 
\epsfig{file=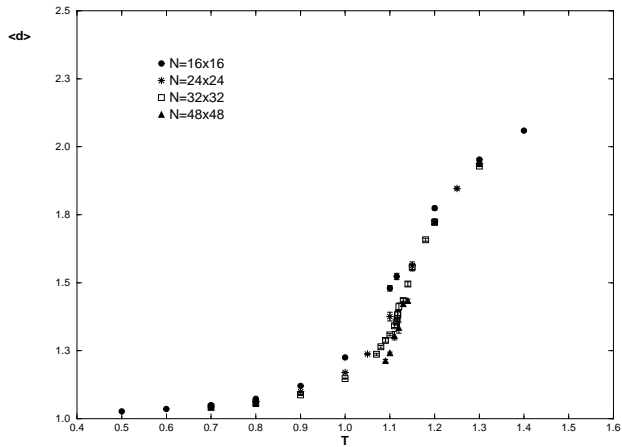,width=6.cm,angle=-90}
\vskip 0.5 truecm
\caption{\small{Averaged NN distance $<d>$ versus $T$,
for $N=16\times 16$, 24$\times$ 24, $32\times 32$ and $48\times 48$}}
\label{dis}
\end{figure}

\begin{figure}[htb] 
\epsfig{file=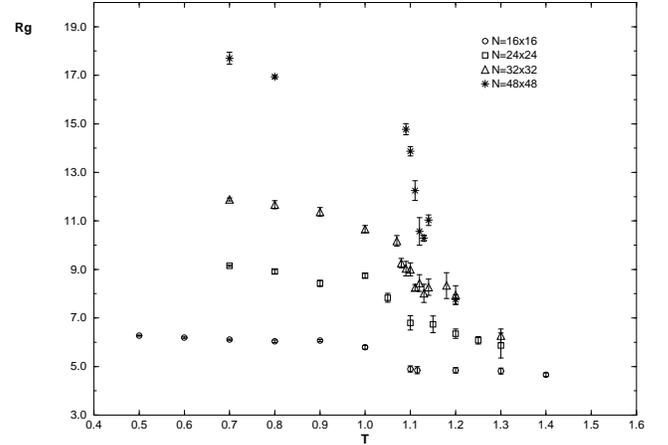,width=6.cm,angle=-90}
\vskip 0.5 truecm
\caption{\small{Averaged gyration radius $R_g$ versus $T$,
for $N=16\times 16$, $24\times 24$, $32\times 32$ and $48\times 48$}}
\label{rg}
\end{figure}

As a first point, we measured $\langle E \rangle$, $<n>$,
$<d>$ and $R_g$ {\it versus}
temperature for sizes $N=16 \times 16$, $N=24 \times 24$,
$N=32 \times 32$ and $N=48 \times 48$.
Results are shown in  figures \ref{ener},\ref{norm},\ref{dis} and \ref{rg}. 
It turned out that equilibration times 
are very 
large for this system. So, as we were interested in critical
properties, we concentrated
our work on the temperature region around the phase transition,
especially for
$N=48 \times 48$.  Figures \ref{ener},\ref{norm},\ref{dis}
and \ref{rg} clearly show a phase
transition between a flat and a crumpled phase at
$T \simeq 1.1$:
there is a sharp jump of $\langle E \rangle $ from the low-$T$
flat phase to  the high-$T$
crumpled phase. Note that the slope of $\langle E \rangle $ increases
with size in the transition region.  The order parameter
 $<n>$ 
drops from a finite value (this value would clearly tend to 1
as $T$ tends to zero)  to
a vanishing value, as expected for an order-disorder transition.
 
At the transition, $R_g$ falls from a finite 
value dependent of the linear size $L$ to a small value
more or less independent of $L$.
This corresponds to the scaling $R_{g} \sim L^{\nu}$, with $\nu =1$
in the flat phase and
$\nu =0$ (indeed, a logarithmic dependence) in the crumpled phase.
It should be noticed that the
NN distance remains finite in the crumpled phase, which means that
monomers actually still form a tethered
membrane even in the
high temperature phase as discussed earlier.
The second purpose of our work was to determine whether the transition
is continuous or not.
A standard method, when using numerical simulations, consists in a
finite size scaling analysis
of the maximum of the specific heat $C_v$.
For a second-order phase transition,
it is expected to grow as the linear system size. For a
discontinuous transition, there are in principle discontinuities
in thermodynamic quantities.
However, for 
small systems, a discontinuous phase transition can appear to be
continuous if the correlation
length is greater than the linear size of the system. In that case,
$C_v^{max}$ is expected to grow 
as the size of the system.\cite{challa} 

We measured $C_{v}^{max}$ for $N=16 \times 16$,
$N=24 \times 24$ and
$N=32 \times 32$ using the histogram technique.\cite{Ferrenberg}
For these small sizes, energy histograms have a single peak
for all temperatures we explored around the transition
(multihistogram method).  They
are found more or
less gaussian.

\begin{figure}[htb] 
\epsfig{file=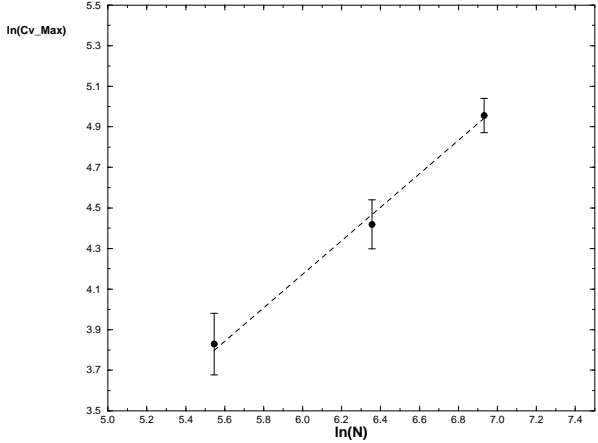,width=6.cm,angle=-90}
\vskip 0.5 truecm
\caption{\small{$C_{v}^{max}$ {\it versus} $N$ in logarithmic
coordinates, for $N=16 \times 16$,
$N=24 \times 24$ and $N=32 \times 32$. The dashed line is the best fit.}}
\label{cvmax}
\end{figure}

In figure \ref{cvmax}, we plot $C_{v}^{max}$
{\it versus} $N$ in logarithmic
scale. Fitting these data, we obtain $C_{v}^{max} \sim N^{x}$ with
$x=0.83(12)$. 
This value is far from 
the value $x=0.5$ expected for a continuous transition. 
It is closer to 1, the theoretical value for
a first-oder transition as discussed above. At this stage, 
in view of this, we conjecture that
the transition is of first order. As seen below this conjecture is confirmed
by histograms made for a larger size.

\begin{figure}[htb] 
\epsfig{file=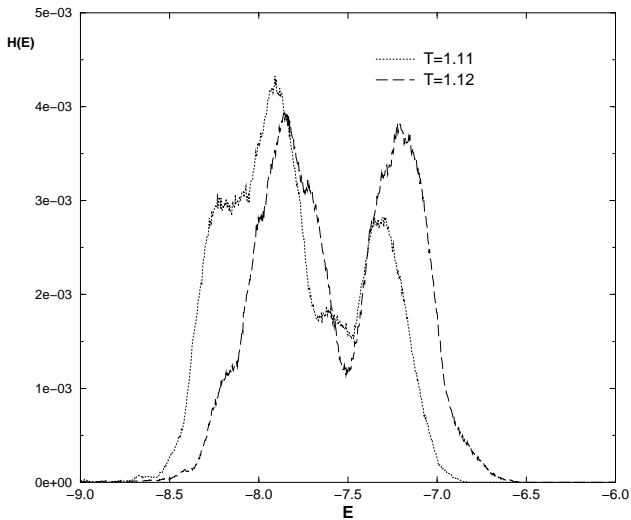,width=7.cm,angle=-90}
\vskip 0.5 truecm
\caption{\small{Normalized histogram for the energy, for $N=48\times 48$.
Double-peak structure showing the first-order character of the phase
transition}}
\label{histo}
\end{figure}

In order to check further the first-order character of the transition, we
increase $L$. For large $L$, if the transition is
of first order, the energy histogram should show
a structure of multiple peaks corresponding to the coexistence
of ordered and disordered phases at the transition.
The system would go back and forth
between these phases resulting in a double-peak energy histogram.
Taking $L=48$, we indeed observed this 
double-peak histogram in the region $T \simeq 1.1$, as shown in
figure \ref{histo}.  This is a
very  strong signal which confirms the first-order character of  
the crumpling transition found earlier by finite-size scaling of $C_v^{max}$.

\section{Conclusion}

We have studied the crumpling phase transition by MC simulation of
a model of tethered membranes with LJ potential energy and
bending rigidity. We have shown clearly the 
first-order nature of the phase transition between
flat and crumpled
phases, in contrast with earlier simulations using gaussian
tethering interaction.  
Note that these early MC results
\cite{kant86,kant87,bowi96,whea93,whea96,baig89,baig94,ambj89,renk90,harn91,espr96}
and also analytical calculations
\cite{davi88,ledo92} show a {\it continuous} phase transition
for models which do not include anharmonic excitations.
We believe that 
the anharmonic nature of the LJ potential used in our model
to some extend contributes to the first-order
transition observed here.  We did not vary in this work the value
of bending rigidity $K$. 
Let us mention however that for $K=0$, there is
no flat phase, the membrane is crumpled at all $T$.

It would be interesting to include 
self-avoidance between non nearest-neighbors in our model.
However, it seems that in that case the precise form of NN potential
is irrelevant: for repulsive self-avoidance
between non nearest-neighbors the membrane is always flat regardless of 
the form of the
potential between NN, even in the
absence of bending rigidity.\cite{kant86,kant87,abraham}
This is interpreted as an effective bending rigidity induced
by excluded volume effect. This can be overcome by an
attractive interaction
between non nearest-neighbors in addition to the
repulsive self-avoiding interaction, leading to
the folding of the membrane at low $T$ and 
a flat phase at high $T$.\cite{abraham2}  Including such repulsive
and attractive interactions 
between non nearest-neighbors in our model is a formidable task which 
is left for future investigations.

\end{document}